\documentclass[fleqn,usenatbib]{mnras}
\usepackage{graphics,epsf}
\usepackage{amsmath}                
\usepackage{amsfonts}               
\usepackage{amssymb}                
\usepackage{epsfig}                 
\usepackage{graphicx}
\usepackage{float}

\newcommand{\G}{{~\rm G}}

\newcommand{\yr}{{~\rm yr}}

\newcommand{\AU}{{~\rm AU}}

\title[GEE toward SN IIB]{Grazing envelope evolution toward Type IIb supernovae}

\author[Soker, N.]{Noam Soker$^{1}$\thanks{E-mail: \href{mailto:soker@physics.technion.ac.il}{soker@physics.technion.ac.il}}
\\
$^{1}$Department of Physics, Technion, Haifa 3200003, Israel\\
\\
}

\begin{document}
\label{firstpage}
\pagerange{\pageref{firstpage}--\pageref{lastpage}}
\maketitle

\begin{abstract}
I propose a scenario where the majority of the progenitors of type IIb supernovae (SNe IIb) lose most of their hydrogen-rich envelope during a grazing envelope evolution (GEE). In the GEE the orbital radius of the binary system is about equal to the radius of the giant star, and the more compact companion accretes mass through an accretion disk. The accretion disk is assumed to launch two opposite jets that efficiently remove gas from the envelope along the orbit of the companion. The efficient envelope removal by jets prevents the binary system from entering a common envelope evolution, at least for part of the time. The GEE might be continuous or intermittent. I crudely estimate the total GEE time period to be in the range of about hundreds of years, for a continuous GEE, and up to few tens of thousands of years for intermittent GEE.
The key new point is that the removal of envelope gas by jets during the GEE prevents the system from entering a common envelope evolution, and by that substantially increases the volume of the stellar binary parameter space that leads to SNe IIb, both to lower secondary masses and to closer orbital separations.
\end{abstract}

\begin{keywords}
stars: jets — stars: supernovae: general — binaries: close — accretion disks
\end{keywords}

\section{INTRODUCTION}
\label{sec:intro}
The goal of this paper is to encourage the consideration of the grazing envelope evolution (GEE) as a major evolutionary phase in the formation of progenitors of type IIb supernovae (SN IIb). I point to some similarities between SN IIb progenitors and post-asymptotic giant branch (AGB) stars that reside in binary systems with an orbital separation of $\approx 1 \AU$. Traditional evolutionary routes predict either wider or closer post-AGB binaries, but do not predict these post-AGB intermediate binaries (post-AGBIBs). I suggested in the past that the GEE can explain post-AGBIBs. I now apply the GEE to progenitors of SN IIb.

\subsection{The grazing envelope evolution (GEE)}
\label{subsec:GEE}
In the traditional treatment of the common envelope evolution (CEE) where a compact star enters the envelope of a giant star, the gravitational energy released by the in-spiraling binary system unbinds the extended envelope of the giant star. However, numerical studies of the CEE that consider only this orbital energy do not manage to eject the common envelope in a persistent and consistent manner (e.g., \citealt{TaamRicker2010, DeMarcoetal2011, Passyetal2012, RickerTaam2012, Nandezetal2014, Ohlmannetal2016, Staffetal2016MN8, NandezIvanova2016, Kuruwitaetal2016, IvanovaNandez2016, Iaconietal2017, DeMarcoIzzard2017, Galavizetal2017}). These simulations might hint on the need for an extra energy source to eject the envelope.

One possible extra energy source is the recombination energy of hydrogen and helium (e.g., \citealt{Nandezetal2015, IvanovaNandez2016}; earlier references therein). However, the efficiency by which the recombination energy can be utilized might be very small \citep{SokerHarpaz2003}, as the opacity in the recombined region is low, and radiation can carry a large fraction of the recombination energy \citep{Harpaz1998}. More relevant to the present paper is that the recombination energy, even if used in full, cannot explain the CEE of massive stars.
Another extra energy source is the accretion process onto the companion (more compact) star, in particular if a large fraction of this energy is carried by jets.
\cite{ArmitageLivio2000} and \cite{Chevalier2012} discussed the ejection of the common envelope by jets that are launched from a neutron star companion. They, however, did not consider this process to be a general CEE mechanism.

I take the view that in many cases jets supply the extra energy to remove the common envelope \citep{Soker2004}, in particular when the companion is a main sequence star (e.g., \citealt{Soker2016Rev}). Jets might be launched even when the accreted gas does not possess enough specific angular momentum to form a fully developed accretion disk \citep{Shiberetal2016, SchreierSoker2016}. From their study of momenta in bipolar planetary nebulae, \cite{BlackmanLucchini2014} suggest that the binary companion might indeed launch energetic jets during the CEE.

The jets not only remove the common envelope, but they also facilitate the accretion onto the companion. Without jets the accretion rate would be much lower, because a high pressure region is built up near the companion (e.g. \citealt{RickerTaam2012, MacLeodRamirezRuiz2015}). The jets remove high-entropy gas, as well as angular momentum and energy, from the vicinity of the accreting companion, hence reducing the pressure around the companion \citep{Shiberetal2016, Staffetal2016MN}.

In the GEE a stellar companion performs a grazing orbit around the envelope of a giant star, accretes mass from its envelope through an accretion disk, and launches jets \citep{SabachSoker2015, Soker2015, Shiberetal2017, ShiberSoker2017}. The GEE takes place when the jets efficiently remove the envelope outside the orbit of the companion, hence preventing a full CEE. The orbital separation might decrease as the GEE takes place, even down to very small separations. In other cases the orbital separation can decrease by a small fraction, or might even increase. In those cases the jets do not manage to eject the entire giant envelope, and the primary star maintains its giant phase.

{{{ It is important to emphasize the differences between the GEE and the case of a Roche lobe over flow (RLOF).(1) In RLOF mass is removed from the giant envelope by the gravity of the companion and by winds. Therefore, the RLOF by itself cannot prevent the system from entering a CEE if the giant expand further or if the system losses synchronization. In the GEE there is an extra energy source, the jets launched by the companion. The jets prevent a CEE, hence substantially increasing the parameter space for the formation of SN IIb. (2) In RLOF mass flows through the first Lagrangian point. Mass transfer in the GEE is a combination of a RLOF and a Bondi-Hoyle-Lyttleton type accretion. (3) In the RLOF the orbital separation is larger than the radius of the giant. In the GEE the companion grazes the envelope at the momentarily contact point, but on average the orbital separation is somewhat smaller than the average radius of the giant. The binary system is in a constant state of `just entering a CEE'.   }}}

Here I propose that the GEE might explain the progenitors of most SN IIb that have giant dimensions and have a stellar companion in an intermediate separation.

\subsection{Post-AGB intermediate binaries (AGBIBs)}
\label{subsec:postAGB}
Traditional calculations of binary stellar evolution that lead to the formation of a post-AGB stars and a main sequence companion include tidal interaction, mass loss, mass transfer, and the CEE. They do not include jets launched by the binary companion that facilitate the removal of the envelope of the AGB progenitor. These calculations predict one of two outcomes of the final orbital separation (e.g., \citealt{Nieetal2012}). Either mass loss causes the orbital separation to increase, such that the final orbital separation is larger than the radius of the AGB progenitor of the post-AGB star, i.e., a wide binary with $a_f \gg 1 \AU$, or the binary system experiences a CEE to form a close binary system with a final orbital separation of $a_f \ll 1 \AU$.
However, there are many intermediate binary systems of a post-AGB star and a main sequence star with an orbital separation just inside the traditional gap (e.g., \citealt{Gorlovaetal2014, Manicketal2017}).

The presence of post-AGBIBs, those with orbital periods in a range where traditional calculations predict that there should be no binaries, calls for the inclusion of a process that was not considered by traditional calculations.
In earlier papers I suggested that this process is the GEE \citep{Soker2015, Soker2016a}.
Indeed, in many, {{{ and probably most \citep{VanWinckel2017b}, }}} of these post-AGBIBs there are indications for jets launched by the more compact companion. As well, in most of them there is a circumbinary disk that testifies to a strong binary interaction, such as that expected when the compact companion grazes the giant envelope.

The central binary star of the Red Rectangle bipolar nebula is a post-AGBIB (e.g., \citealt{VanWinckel2014} for its properties). The main sequence companion to the post-AGB star launches jets \citep{Wittetal2009} with wide opening angles \citep{Thomasetal2013}.
\cite{Gorlovaetal2012} find the post-AGBIB system {BD$+46^\circ442$} to launches jets, and  \cite{Gorlovaetal2015} report a bipolar outflow from the post-AGBIB system {IRAS~19135+3937}.

The GEE explanation for post-AGBIBs posits that the system experiences a relatively short, tens to hundreds of years, GEE phase during the AGB phase of the primary star. After the intense removal of envelope mass by jets, the AGB envelope contracts and the star evolves toward the post-AGB phase. Mass transfer and jets launching continue beyond the GEE.
In the preset study I propose that progenitors of SN IIb experience a GEE phase similar in some aspects to the GEE phase of post-AGBIBs.

An analogue between massive stars and post AGBIBs was made in the past. \cite{Kastneretal2010} and \cite{VanWinckel2017} compare the circumbinary disks of post-AGBIBs with those of B[e] supergiants. The present study then might lead to the conclusion that some B[e] supergiants evolve to become SN IIb.

\subsection{Type IIb supernovae (SNe IIB)}
\label{subsec:SNIIb}
SNe IIb are core collapse supernovae (CCSNe) that have strong hydrogen lines at early times, days after explosion, which later become weaker or even disappear. This implies that the CCSN progenitor star has a very little hydrogen mass at the time of explosion,
$M_{\rm H} \simeq 0.03-0.5 M_\odot$ (e.g., \citealt{Woosleyetal1994, Meynetetal2015, Yoonetal2017}).
SNe IIb amount to about 10.3 per cents of all CCSNe \citep{Smithetal2011, Shivversetal2017}, and their relative rates do not seem to depend on the mass of their host galaxies \citep{Grauretal2017b}.

\cite{Alderingetal1994} find that the photometry of SN 1993J is inconsistent with a single star, and it was most likely a binary system, as suggested by \cite{Podsiadlowskietal1993}.
\cite{Foxetal2014} mention that the binary scenario for SN 1993J is also supported by the presence of a flattened CSM \citep{Mathesonetal2000} that was formed by a very high mass loss rate from the progenitor. Mass loss in a flat disk (ring) is one property that I take to connect SN IIb progenitors to post-AGBIBs.

\cite{Kilpatricketal2017} estimate that the radius of the progenitor of SN 2016gkg at explosion was $R=260 R_\odot$. Their best fitting binary evolution model is for an initial primary and secondary masses of $M_{1i}=15M_\odot$ and $M_{2i}=1.5 M_\odot$, respectively, a final primary mass of $M_{1f}=5.2 M_\odot$, and an initial orbital period of $P_i= 1000~$days.
The presence of a giant star despite an enhanced mass loss rate induced by a binary companion is another property that I take to relate SN IIb progenitors to post-AGBIBs.

Several studies examine the formation of SN IIb in a more systematic search of the parameter space. \cite{Claeys2011} expand the earlier work of \cite{StancliffeEldridge2009} and study evolutionary routes that might lead to SN IIb. They find that under their assumptions binary evolution predicts about 0.6 per cents of all CCSNe to be SN IIb.
To increase this rate they find that they should consider low accretion efficiencies by the companion, in combination with limited angular momentum loss from the system.
{{{ \cite{OuchiMaeda2017} also find that a large fraction of the mass lost by the giant star should be lost from the system. }}}
In the present study I attribute these to removal of mass by jets launched by the companion. The jets both reduce mass accretion onto the companion, and remove mass from the primary stellar progenitor with relatively low specific angular momentum.

In the most recent study I am aware of, \cite{Sravanetal2017} find that single star evolution cannot explain the progenitors of SNe IIb (also \citealt{Sravan2016}). They further find that their binary evolutionary routes account for about only 10 per cents of all SN IIb. The last finding hints on the possibility that non-traditional evolutionary routes that include a new ingredient play a major role in leading to SN IIb. In the present study I propose that this new ingredient is the GEE. If the majority of SNe Ib and SNe Ic also result from binary systems (e.g., \citealt{Grauretal2017a}), then the GEE might play a role in their formation as well.

\section{TIME SCALES}
\label{sec:time}
There are three issues related to the GEE.
The first issue is the energy budget. Since the secondary star here is massive as well (e.g. \citealt{Claeys2011, Yoonetal2017}), by accreting even a small fraction of its own mass
($M_{\rm acc}/M_2 \approx 0.1-0.3$) through an accretion disk, it can launch jets that carry enough energy to unbind the envelope of the primary star. I will not repeat the energy calculation here,  as more details can be found in an earlier paper \citep{Soker2015}.
The second issue is the distribution of angular momentum during the GEE. I postpone this treatment to section \ref{sec:AM}.

I here consider the relevant time scales that teach us on the behavior of the binary system during the GEE. Following the first paper on the GEE \citep{Soker2015},
I consider the Kelvin-Helmholtz (thermal) time-scales $\tau_{\rm KH-env}$, and the tidal time scale $\tau_{\rm T-ev}$. In this exploratory study I only crudely estimate the timescales, to show the feasibility of the propose GEE scenario for SN IIb progenitors.

For the onset of the GEE, I consider the response of the outer envelope. The core and inner envelope layer do not change much during the initial phase of the GEE. Crudely then, the envelope initial thermal time scale is given by
\begin{eqnarray}
\tau_{\rm KH-env} = \beta_{\rm env} \frac { \G M_{1i} M_{\rm env} }{ R_{1i} L_1 } \approx 100
\left( \frac{M_{1i}}{15 M_\odot} \right)
\left( \frac{M_{\rm env}}{5 M_\odot} \right) \nonumber \\
\times
 \left(\frac{L}{10^5 L_\odot} \right)^{-1} \left( \frac{R_{1i}}{4 \AU}
\right)^{-1} \yr.
  \label{eq:tkh2}
\end{eqnarray}
where the increase of density inward was taken into account with $\beta_{\rm env} \simeq 2-8$ (see \citealt{Soker2008} for the calculation) in deriving this crude value.

The scaling of the primary luminosity $L_1$, radius $R_1$ and mass $M_1$ is based on studies of SN IIb progenitors (e.g., \cite{Claeys2011, Yoonetal2017}.

It should be noted that the jets in the GEE remove a substantial amount of mass. After substantial mass removal massive stars expand and develop an intense wind (e.g., \citealt{Kashietal2016}). Although the photosphere expands, the dense part of the envelope contracts a little \citep{Kashietal2016}. On a thermal time scale the entire envelope expands.
Later on, with further mass removal, the envelope and photosphere contract.
 Namely, the GEE might take place when the primary radius is larger than its radius at explosion.

The tidal spiral-in timescale is taken from \cite{Soker1996} that uses expressions derived by   \cite{Zahn1977} and \cite{VerbuntPhinney1995}, and it reads
\begin{eqnarray}
\tau_{\rm T-ev} \simeq 60  
  \left(   \frac {a}{1.2R_1} \right)^{8}
   \left( \frac {L} {10^5 L_\odot} \right)^ {-1/3} \nonumber \\
      \times
   \left( \frac {R_1}{4 \AU} \right)^ {2/3}
  \left( \frac{M_{\rm {env}}}{0.33M_1} \right)^{-1}
   \left( \frac{M_{\rm {env}}}{5M_\odot} \right)^ {1/3} \nonumber \\
      \times
   \left( \frac{M_2}{0.2M_1} \right)^ {-1}
   \left( \frac{\Omega_{\rm orb} -\omega_{1}}{0.1 \Omega_{\rm
   orb}}\right)^{-1} \yr,
     \label{eq:tidal}
\end{eqnarray}
where $a$ is the orbital separation, $\omega_{1}$ is the rotational angular velocity of the primary envelope, and $\Omega_{\rm orb}$ is the orbital angular velocity.

The evolution proceeds as follows. As the companion grazes the envelope of the giant primary, it launches jets that efficiently remove the envelope in the vicinity of the companion. The removal of mass from the outer envelope layers increases the spiraling-in time given by equation (\ref{eq:tidal}), both because of the mass loss itself and because of weakening of the tidal interaction. The companion spirals-in slowly, while the envelope expands on a thermal time scale given by equation (\ref{eq:tkh2}). The system is now further away from synchronization, namely,
$\omega_{1} < 0.9  \Omega_{\rm orb}$. This reduces the spiraling-in time. On the other hand, most of the envelope mass is deep inside the orbital radius, hence $R_1 < a/1.2$. This increases the spiraling-in time. Over all, the spiraling-in time scale is from about tens of years to several hundreds years at early phases of the GEE. Later on, when the mass of the envelope is reduced and the envelope starts to contract, the time scale increases, possibly up to thousands of years.

It is quite possible that in some cases the jets remove lots of mass in few orbits \citep{Shiberetal2017}, and mass from the inner envelope does not keep in pace with mass removal to fill for the lost envelope; The launching of jets ceases for a while. As the envelope later expands on a thermal times scale, the process repeats itself. In this scenario the intermittent GEE might last for a longer time of $\approx 10^4 \yr$.

Over all, the GEE takes place over a times scale of $\approx 10^3$ (for continuous GEE) to $\approx {\rm few} \times 10^4 \yr$ (for an intermittent GEE). The binary system can then continue to evolve as a detached binary system, or the companion might accrete from the wind of the primary star, until explosion.
The key point is that the removal of envelope mass by jets prevents the system from entering a full-scale CEE, and by that substantially increases the volume of the parameter space that leads to SN IIb, both toward lower secondary masses and smaller orbital separations. In other words, part of the parameter space that is designated as contact binaries by \cite{Claeys2011}, might now lead to SN IIb that has an extended envelope at explosion.

\section{ANGULAR MOMENTUM}
\label{sec:AM}
Three factors make the GEE different from some other evolutionary routes.

(1) Unlike in the CEE, energy is not a consideration for determining the final orbital separation, as the accretion onto the secondary star provides most of the required energy to remove the envelope.

(2) The jets remove envelope gas with a specific angular momentum smaller than that of the secondary star. For example, mass that is lost through the second Lagrangian point (beyond the secondary star) leaves the system with high value of specific angular momentum $j_{\rm L2}$. In the GEE most of the mass that is expelled by the jets have a relatively low value of specific angular momentum, $j_{\rm out} \ll j_{\rm L2}$.

(3) Because of substantial envelope mass removal, the secondary star does not grow much in mass.
Therefore, although the secondary accretes some mass, this mass is very small compared with the envelope mass, and I neglect the accreted mass in the expression for the angular momentum carried by the envelope

I apply these to show that the system can avoid a CEE, and the orbital separation can stay large. I take the initial and final mass of the primary and secondary stars to be $M_{1i}$, $ M_{1f}$, $M_{2i}$, $ M_{2f}$, respectively. Let the GEE start at an orbital separation $a_i$, and end at an orbital separation $a_f$.

I assume that when the GEE starts the orbital separation is about equal to the primary radius, $a\simeq R_1$, and that the system is close to synchronization $\Omega_{\rm orb} \simeq \omega_{1}$. The moment of inertia of the envelope is $I_{\rm env} = \eta_1 M_{\rm env} R_1^2$, where $M_{\rm env}= M_{1i} - M_{1f}$ is the envelope mass, and $\eta_1 \simeq 0.1-0.2$. The initial angular momentum of the envelop is, under the above assumptions, $J_{\rm env,i}=\eta_1 M_{\rm env} a^2_i \Omega_{\rm orb}$, with $\Omega_{\rm orb}=(GM_i/a_i^3)^{1/2}$, and $M_i = M_{1i}+M_{2i}$ is the total initial mass. Any deviation from initial synchronization can be absorbed in what follows inside $\eta_1$.

 The initial angular momentum of the system is therefore
\begin{equation}
J_i= \left(  GM_ia_i  \right)^{1/2} \left( \frac{M_{1i}M_{2i}}{M_i} + \eta_1 M_{\rm env} \right).
     \label{eq:J0}
\end{equation}

The average specific angular momentum carried by the envelope out of the system is expressed with a parameter $\epsilon_j$, and reads,
\begin{equation}
J_{\rm env,out} = M_{\rm env} \epsilon_j j_{2i},
     \label{eq:Jenv}
\end{equation}
where $j_{2i}=(GM_i a_i)^{1/2} (M_{1i}/M_i)^2$ is the initial specific angular momentum of the secondary star around the center of mass.

After mass loss starts, the secondary does not maintain the envelope in synchronization any more. So the specific angular momentum on the surface of the giant becomes much smaller than $j_2$.
As well, as more mass is lost from the system the specific angular momentum of the secondary star decreases, $j_2 < j_{2i}$. Although the jets carry specific angular momentum equals to that of the secondary star at a given time, the jets do not carry large amount of mass. It turns out that on average along the entire mass loss process $J_{\rm env,out} < j_{2i}$.
I crudely estimate that $\epsilon_j \simeq 0.2-0.4$, and I will scale it with $\epsilon_j =0.3$.

The final angular momentum of the binary system is
\begin{equation}
J_f= \left(  GM_f a_f  \right)^{1/2}  \frac{M_{1f}M_{2f}}{M_f},
     \label{eq:Jf}
\end{equation}
where $M_f = M_{1f}+M_{2f}$ is the total final mass of the binary system at the end of the GEE.
Substituting equations (\ref{eq:Jenv})-(\ref{eq:Jf}) in the condition of angular momentum conservation $J_f=J_i-J_{\rm env,out}$, yields the ratio of the final orbital separation to the initial orbital separation
\begin{eqnarray}
\frac{a_f}{a_i}  \simeq
\left( \frac{ M_{1i} }{ M_{1f} } \right)^2  \left( \frac{ M_{f} }{ M_{i} }  \right)
\left( \frac{ M_{2i} }{ M_{2f} }  \right)^2  \nonumber \\
\times
\left[  1+ \left(  \eta_1 \frac{M_i}{M_{1i}} - \epsilon_j \frac{M_{1i}} {M_i}      \right)
\frac{ M_{\rm env} }{ M_{2i} } \right]^2
     \label{eq:af}
\end{eqnarray}

As an example I take a case with a low mass secondary star. Traditional calculations that do not include jets launched by the secondary star end with a CEE (e.g., \citealt{Claeys2011}). This can lead to a CCSN similar to 1987A, but not to a SN IIb. When jets are considered, the system experiences the GEE, and avoids a CEE. I use these physical values. Initial and final primary masses, $M_{1i}=15 M_\odot$, $M_{1f}=5 M_\odot$, respectively, initial and final secondary masses $M_{2i}=2M_\odot$, $M_{2i}=2.5 M_\odot$, respectively, $\eta_1=0.15$, and $\epsilon_j=0.3$. For these values $M_{\rm env}=10M_\odot$, $M_i=17M_\odot$, and $M_f=7.5M_\odot$.
Substituting these values in equation (\ref{eq:af}) gives $a_f \simeq 0.7 a_i$. For $a_i=4 \AU$ when GEE starts, the final orbital separation is $a_f \simeq 2.8 \AU$ for these parameters. At explosion, the primary star can be a blue supergiant, well inside the orbit of the secondary star.
For other parameters the orbital separation might increase, and the evolution might take place via an intermittent GEE.

The following points should be considered.
(1) Equation (\ref{eq:af}) gives the final orbital separation based only on angular momentum conservation. However, some fraction of the orbital energy is expected to be channelled to mass removal, so the final orbital separation might be smaller that that given by equation (\ref{eq:af}), but still much larger than that predicted by the traditional CEE.
(2) If the final orbital separation given by equation (\ref{eq:af}) is much larger than the initial primary radius, then it overestimates the value of $a_f$. Simply, if $a_f$ is much larger than $R_1$, then even if the secondary launches jets, they will not remove envelope mass.

The point to take from equation (\ref{eq:af}) is that the GEE allows the primary to explode as a giant, but with little hydrogen-rich envelope mass.

\section{SUMMARY}
\label{sec:summary}
Motivated by the limitation of traditional binary calculations to account for the observed number of SNe IIb (section \ref{subsec:SNIIb}), and based on some similarities of post-AGBIBs (section \ref{subsec:postAGB}) to SN IIb progenitors, I propose that the majority of SN IIb progenitors are formed via the GEE (section \ref{subsec:GEE}).
The key ingredient of the GEE is that jets launched by the secondary star supply most of the energy to remove the envelope of the primary star. This prevents the binary system from entering a CEE, and leaves the orbital separation very large, hence allowing the SN IIb progenitor to explode as a giant star. The binary interaction ensures the removal of most of the hydrogen-rich envelope.

In section \ref{sec:time} I crudely estimated that the GEE can last from several hundreds of years, for a continuous GEE, to over ten thousands years, for an intermittent GEE.

In section \ref{sec:AM} I derived an expression for the final orbital separation under the assumption that it is determined by angular momentum conservation.
The final orbital separation in equation (\ref{eq:af}) is sensitive to the value of $\epsilon_j$ and $M_{\rm env}/M_{2i}$. The equation cannot be used if it gives a final orbital separation that is much larger that the maximum radius the primary star achieves along its evolution. The main new point of this exploratory study is that when the GEE is considered, the binary system might avoid a CEE for a much larger volume of the parameter space than that allowed by traditional calculations. The evolution will be similar to the proposed GEE explanation to post-AGBIBs.

The next step is to perform population synthesis studies of the GEE. These will tell us whether evolutionary routes that include the GEE can account for most SN IIb.


\label{lastpage}
\end{document}